# Raman spectroscopy study of the interface structure in $(CaCuO_2)_n/(SrTiO_3)_m$ superlattices.

D. Di Castro,[1] S. Caramazza,[2] D. Innocenti,[1] G. Balestrino,[1] C. Marini,[3] P. Dore,[4] and P. Postorino[5,a)]

[1]CNR-SPIN and Dipartimento di Ingegneria Civile e Ingegneria Informatica, Università di Roma Tor Vergata, Via del Politecnico 1, 00133 Roma, Italy

[2]Dipartimento di Fisica, Università di Roma Sapienza, Piazzale Aldo Moro 2, 00185 Roma, Italy

[3]European Synchrotron Radiation Facility, BP 220, 38043 Grenoble Cedex, France

[4]CNR-SPIN and Dipartimento di Fisica, Università di Roma Sapienza, Piazzale Aldo Moro 2, 00185 Roma, Italy

[5]CNR-IOM and Dipartimento di Fisica, Università di Roma Sapienza, Piazzale Aldo Moro 2, 00185 Roma, Italy

Abstract

Raman spectra of $CaCuO_2/SrTiO_3$ superlattices show clear spectroscopic marker of two structures formed in $CaCuO_2$ at the interface with $SrTiO_3$. For non-superconducting superlattices, grown in low oxidizing atmosphere, the 425 cm$^{-1}$ frequency of oxygen vibration in $CuO_2$ planes is the same as for CCO films with infinite layer structure (planar Cu-O coordination). For superconducting superlattices grown in highly oxidizing atmosphere, a 60 cm$^{-1}$ frequency shift to lower energy occurs. This is ascribed to a change from planar to pyramidal Cu-O coordination because of oxygen incorporation at the interface. Raman spectroscopy proves to be a powerful tool for interface structure investigation.

______________________________

High-temperature superconductors (HTS) are intrinsically multi-layered materials where a block with "infinite layer" (IL) structure, containing $CuO_2$ planes, is stacked with other metal-oxide layers, called "charge-reservoir" (CR) blocks. When the latter are charge unbalanced by cation substitution or oxygen deficiency/excess, they provide mobile holes or electrons to the $CuO_2$ layers. Such a structural configuration of HTS has led investigators to artificially synthesize, using layer-by-layer deposition techniques, heterostructures based onto two different materials, neither superconducting on its own, acting as IL and CR blocks respectively. Following this line, high $T_c$ superconductivity was reported in heterostructures consisting of alternate layers of insulating (IL) and metallic (CR) cuprates.[1,2,3] More recently, it was found that high $T_c$ superconductivity can occur even if two insulating oxides are used, as in $CaCuO_2/SrTiO_3$ (CCO/STO) superlattices.[4] In this cuprate/titanate heterostructure the CCO is the IL block whereas the role of CR is played by the CCO/STO interface layers. Superconductivity indeed does appear only if the superlattices are grown under strongly oxidizing conditions, due to extra oxygen ions entering at the interface.[4,5,6] The larger the number of these oxygen ions is, the higher the hole doping of the CCO block and the $T_c$ are (a maximum $T_c$ ~40 K has been obtained).[4] Excess oxygen ions occupy the apical position for the Cu ions of the $CuO_2$ planes close to the interface.[4,5,6] Therefore, different structures are present in the CCO layers close to the boundaries with $SrTiO_3$ and in those ones far from them. Indeed, the latter should retain their IL structure, with purely planar Cu-O coordination, whereas pyramidal Cu-O coordination can occur in the former. Recently, the occurrence of similar structural conditions has been also foreseen in films of IL system $ACuO_2$ (A = Ca, Sr, Ba) in a theoretical paper by Zhong et al.[7] and experimentally proven in the case of $SrCuO_2/SrTiO_3$ superlattices.[8] Zhong et al.[7] found that, in order to reduce the electrostatic instability, naturally present in these highly polar systems, the structure changes, below a critical $ACuO_2$ thickness, from the $CuO_2$ planar (IL) to a chain structure by a tilting of the $CuO_2$ planes. This transition is obtained by moving oxygen atoms from the $CuO_2$ planes to the Ca planes, thus leading to the presence of apical oxygen.

Raman spectroscopy demonstrated to be a powerful tool to investigate layered structures[9,10] and to provide a careful characterization of film samples.[11] In particular, in cuprates with IL structure, the effects of the hole doping of $CuO_2$ planes, obtained by either introduction of excess oxygen atoms[12] or excitation through the charge transfer gap,[13] has been studied in detail. In both cases it was found that Raman-forbidden modes related to oxygen vibrations can appear in the Raman spectra. In particular, in pure CCO films, the larger the amount of excess oxygen ions is, the higher the intensities of these Raman spectral features are.[12] As mentioned above, excess oxygen ions not only provide holes to the $CuO_2$ planes, but also modify the oxygen local arrangement, which, in turn, can affect the Raman spectrum.

In this work we report on a study of the interface structure variations in CCO/STO superlattices as a function of the oxidizing power of the growth atmosphere by means of Raman spectroscopy. To this aim we used the Pulsed Laser Deposition technique to synthesize several $(CCO)_n/(STO)_m$ superlattices ($n$ and $m$ are the number of unit cells of CCO and STO, respectively), following the procedure reported in Ref. 4. All the superlattices are grown with almost the identical architecture, i.e. 18-20 repetitions of the $(CCO)_n/(STO)_m$ supercell with $n = 4 \div 5$ and $m = 2 \div 3$. Both $NdGaO_3$ (110) and $LaAlO_3$ (001) substrates were employed.

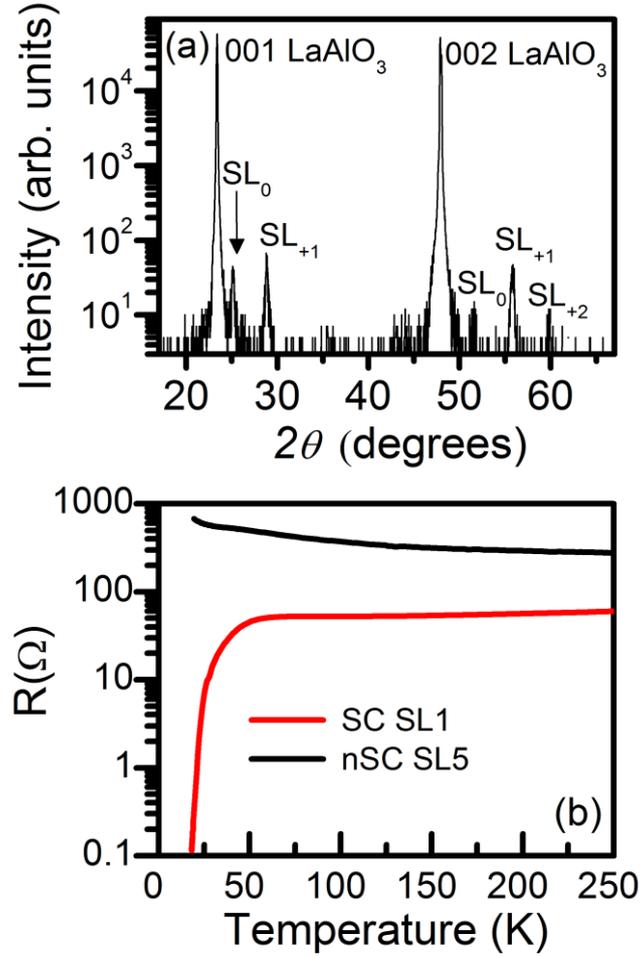

FIG. 1. (a) XRD pattern from the $(CCO)_{4.5}/(STO)_{2.5}$ superlattice (sample SL5) including the (001) and the (002) reflections of the $LaAlO_3$ substrate. $SL_{+i}$ mark the satellite peaks around the average structure peak $SL_0$. (b) R(T) for the non-superconducting SL5 and for the superconducting SL1.

Properly tuning the growing conditions of CCO/STO superlattices, superconducting (SC) and non-superconducting (nSC) samples were grown. For sake of comparison CCO and STO single phase films were also synthesized. The list of all the investigated samples, together with their relevant growing parameters, are reported in Table I. The structural quality of the samples is always very good. As an example, the x-ray diffraction (XRD) pattern of the SL5 sample (see Table I) is shown in Fig. 1(a). The presence of sharp satellite peaks ($SL_{+1}$, $SL_{+2}$) close to those ascribed to the average structure ($SL_0$), indicates the formation of a high quality superlattice with period $\Lambda \approx 24.5$ Å. The temperature dependences of the resistance R for a SC and a nSC sample are shown in Fig. 1(b). We want to stress that, as already reported in Ref. 4, the conductivity of the superlattices slightly varies with *n* and *m*, whereas it substantially increases with increasing the oxidizing power of the growth atmosphere. Under low oxidizing power (pure oxygen) the R(T) exhibits a semiconductor-like behavior (see SL5 in Fig. 1(b)), whereas under highly oxidizing power (oxygen enriched with 12% ozone) the R(T) shows a metallic behavior followed by a superconducting transition on cooling (see SL1 in Fig. 1(b))

TABLE I. List of the investigated samples and their growing parameters. (SC) and (nSC) labels mark superconducting and non-superconducting samples. *n*, *m* are the numbers of unit cells of CCO and STO, respectively. In the last two columns, $O_2$ and $O_2+O_3$ indicate pure oxygen and ozone enriched growth atmosphere, respectively. P is the pressure.

| Sample    | *n*+*m*   | Thick [Å] | Growth [atm] | P [mbar] |
|-----------|-----------|-----------|--------------|----------|
| SL1 (SC)  | 6.5       | 480       | $O_2+O_3$    | 0.8      |
| SL2 (SC)  | 6.5       | 480       | $O_2+O_3$    | 0.8      |
| SL3 (SC)  | 6.5       | 435       | $O_2+O_3$    | 0.8      |
| SL4 (SC)  | 7.0       | 490       | $O_2+O_3$    | 1.1      |
| SL5 (nSC) | 7.0       | 440       | $O_2$        | 0.2      |
| SL6 (nSC) | 7.5       | 460       | $O_2$        | 0.8      |
| Film CCO  | 50 (m=0)  | 160       | $O_2+O_3$    | 0.8      |
| Film STO  | 40 (n=0)  | 160       | $O_2$        | 0.8      |

Raman spectra were measured in back-scattering geometry, using a confocal micro-Raman spectrometer equipped with a He-Ne Laser, a CCD detector and a notch filter to reject the elastic contribution. The microscope, with a 100 X objective and a very small confocal diaphragm (20 μm), provided a laser spot few μm$^2$ wide at the sample surface with a small depth of field, thus significantly reducing the collection of the Raman signal from the substrate.[14] Low-noise Raman spectra were collected with acquisition times ranging from 40 to 60 minutes over the 200÷1000 cm$^{-1}$ frequency range, with a spectral resolution better than 3 cm$^{-1}$. Although we collected Raman spectra from CCO/STO samples grown on both NdGaO$_3$ and LaAlO$_3$ obtaining similar results, here we report only those obtained with LaAlO$_3$ since its simply structured Raman spectrum allows for a better extraction of the Raman response from the superlattice only.[14,15]

Raman spectra collected from different superlattices and the reference spectra of CCO and STO films and of LaAlO$_3$ substrate are shown in Fig. 2. Looking at Fig. 2(b), we notice that the only significant spectral structure of LaAlO$_3$ is the sharp peak at 485 cm$^{-1}$, which can be clearly identified in all the other spectra shown in Fig. 2. The signal from the STO film is hardly detectable and the measured spectrum is actually identical to that from the LaAlO$_3$ substrate. This result, according with previously reported data collected on thin films,[16,17,18] suggests a major relevance of the CCO response in the composite samples. As to the pure CCO film, according to the factor group analysis, all phonons at Γ point are forbidden for the first-order Raman scattering in the infinite layer structure.[12,19] As pointed out by Kan *et al.*,[12] the forbidden phonons are activated by the excess oxygen induced structural disorder, which relaxes the *k*-selection rule for the first-order Raman scattering.[20] The present spectrum of CCO film is in good agreement with previous results[12] and it is characterized by the presence of broad spectral bands around 500-700 cm$^{-1}$ and weak contributions close 400 cm$^{-1}$. Measured spectra of the CCO/STO superlattices, see Fig. 2(a), show structures at about the same frequencies but remarkably different in shape. This suggests an important role of the rearrangement of thin CCO layers in the heterostructure due to the creation of interfaces. Indeed, the specific architecture of our superlattices, with both CCO and STO blocks just few unit cells thick, enhances the contribution of the interface CCO layers with respect to the inner CCO layers far from the interface. Since the spectral contribution from STO is expected to be negligible, in superlattice samples we are actually measuring the Raman spectrum of CCO steadily out of structural equilibrium condition.

In order to single out and analyze the spectral response of the superlattice CCO/STO film only, we fitted the measured spectra by using a standard procedure (see Ref. 15) with a model function including a sum of damped harmonic oscillators (DHO) for the phonon contributions and a background accounting for both the electronic contribution and the residual signal from the $LaAlO_3$ substrate.[10,21,22] This procedure provides an excellent description of all the measured spectra as shown in Fig. 2(a). The phonon components of the CCO/STO superlattices and the CCO film thus obtained are shown in Fig. 3(a). The contribution around 400 cm$^{-1}$ (A), found in all samples, is ascribed to the oxygen bending mode in the $CuO_2$ plane according to Ref. 12. Three (B to D) and four (B to E) DHO contributions have been used to fit the broad band around 500-700 cm$^{-1}$ for superlattices samples and pure CCO film, respectively.

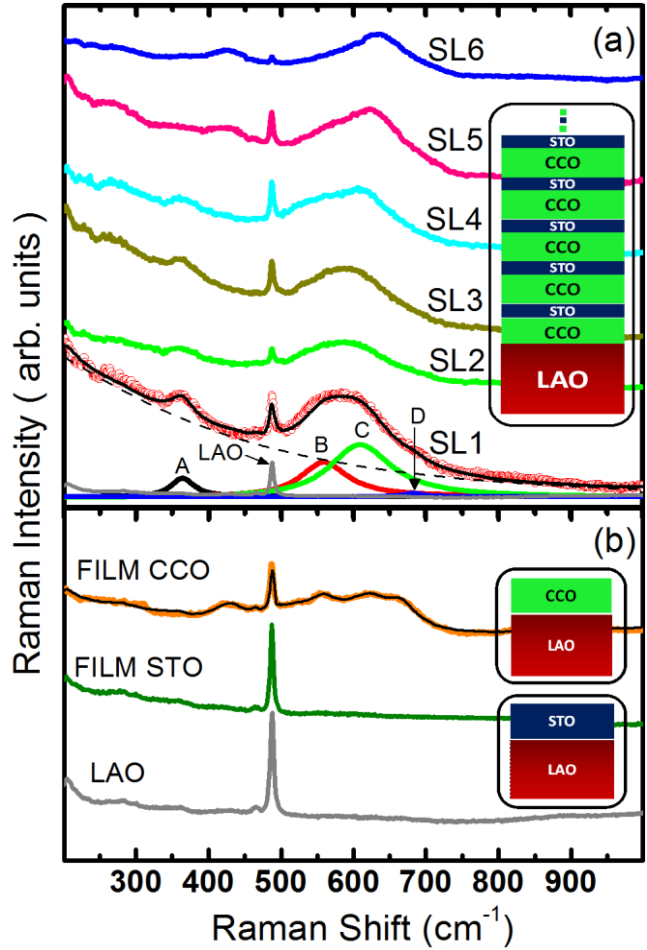

FIG. 2. Raman spectra of the investigated samples. For each kind of sample, a schematic of its structure is reported. (a) Raman spectra of superlattices (CCO)n/(STO)m on LaAlO3 substrate. In the SL1 case, the best profile (full line) is shown. The four dominant DHO components (A, B, C, D), the background contribution (dashed line) and the LaAlO3 spectrum (grey line) are shown separately. (b) Raman spectra of CCO film, STO film, and LaAlO3 substrate. In the CCO case, the best fit profile (full line) is shown.

The best fit peak frequencies of phonon A are shown in Fig. 3(b) for superlattices and CCO film and those ones of C phonon (Cu-O stretching mode[12]) in Fig. 3(c) for superlattices only. Both peaks show a characteristic two steps behavior, as a function of the oxidizing growth conditions, although with a different frequency shift, namely $\Delta\nu \sim 60$ cm$^{-1}$ and $\Delta\nu \sim 30$ cm$^{-1}$ for the A and C phonon, respectively. We will focus our discussion on the well recognizable A-peak since it is well detectable and clearly isolated from other spectral features (see Fig. 3(a)) although quite similar arguments apply also to the frequency of the C-phonon.

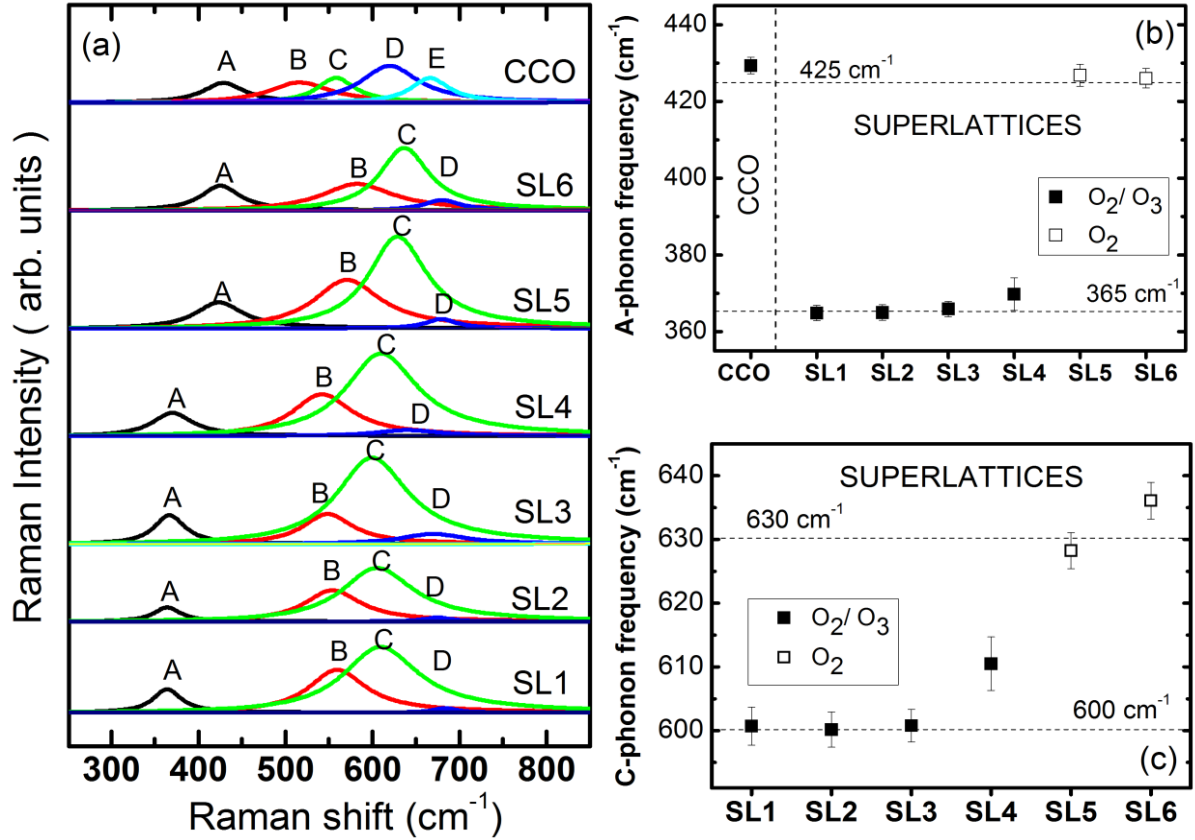

FIG 3. Best fit results: (a) phonon spectral components of superlattices and CCO film; (b) A-phonon frequencies for superlattices and CCO film; (c) C-phonon frequencies for superlattices. Open and full symbols refer to pure oxygen and ozone enriched growth atmosphere respectively.

On the basis of the literature results previously discussed[4,7,8] we can argue that the two different frequencies observed for the A-phonon ($A_{High}$ = 425 cm$^{-1}$, $A_{Low}$ = 365 cm$^{-1}$) could be related to the occurrence of the two possible structures with planar or pyramidal oxygen coordination of CCO at the interface with STO. The $A_{High}$ value is indeed measured in the case of nSC superlattices as well as in pure CCO film (in agreement with Ref. 12) and thus ascribed to the infinite layer structure (planar coordination). The $A_{Low}$ value is on the other hand ascribed to the pyramidal coordination which occurs only in SC superlattices where the strongly oxidizing growth conditions allows extra oxygen ions enter the Ca or Sr planes at the interface[4,5] in the apical position for Cu.

The above assignment is also coherent with the paper by Zhong *et al.*,[7] where the prediction of a transition from planar to chain structure (oxygen in apical position for Cu) is reported for CCO when the thickness of the film is reduced down to ~2 unit cell. Since all our superlattices have CCO block larger than 2 unit cell, we expect the CCO blocks to have planar coordination and the A-phonon at $A_{High}$. This is indeed obtained for all nSC superlattices, whereas the A-phonon at $A_{Low}$ found in SC superlattices suggests that the use of a strongly oxidizing atmosphere drives a transition towards the pyramidal structure of the CCO blocks by the introduction of extra oxygen allocated in apical position along the border region.

We note here that the pure CCO film, although grown in a highly oxidizing atmosphere (as the SC superlattices), shows the peak A in the position $A_{High}$ (see Fig. 3(b)). This finding leads to conclude that the amount of extra apical oxygen ions which is possible to force within the bulk of CCO structure is not enough to be visible with Raman spectroscopy and to obtain a significant hole doping of the system. The interface between CCO and STO thus open an *easy way* to allocate extra oxygen atoms in apical positions, which also provide holes to the $CuO_2$ planes, necessary for superconductivity.[4]

In conclusion, we studied Raman spectra of high quality epitaxial films of CCO/STO superlattices deposited on $LaAlO_3$ substrate. A careful data analysis allowed identifying a clear spectroscopic marker of the two possible oxygen arrangements around the Cu atoms at the interface. Both the bending and stretching modes of the oxygen atoms indeed show high-frequency values when the planar Cu-O coordination (infinite layer structure) is realized, and low-frequency values when, owing to the highly oxidizing growth atmosphere, a pyramidal structure is obtained. Raman spectroscopy thus proves to be a powerful tool to investigate complex systems with a high sensitivity to interface states, particularly important in the physics of artificial heterostructures. The present results are of particular interest since the different Cu-O spatial coordination is strongly related to the onset of a superconductive phase.

This work was partly supported by the Italian MIUR (Grant No. PRIN-20094W2LAY).

[a] Electronic mail: paolo.postorino@roma1.infn.it

[1] G. Balestrino, S. Martellucci, P. G. Medaglia, A. Paoletti, and G. Petrocelli, Phys. Rev. B **58**, R8925 (1998)

[2] C. Aruta, G. Ghiringhelli, C. Dallera, F. Fracassi, P. G. Medaglia, A. Tebano, N. B. Brookes, L. Braicovich, and G. Balestrino Phys. Rev. B **78**, 205120 (2008).

[3] A. Gozar, G. Logvenov, L. Fitting Kourkoutis, A. T. Bollinger, L. A. Giannuzzi, D. A. Muller, and I. Bozovic, Nature **455**, 782 (2008)

[4] D. Di Castro, M. Salvato, A. Tebano, D. Innocenti, C. Aruta, W. Prellier, O. I. Lebedev, L. Ottaviani, N. B. Brookes, M. Minola, M. Moretti Sala, C. Mazzoli, P. G. Medaglia, G. Ghiringhelli, L. Braicovich, M. Cirillo, and G. Balestrino, Phys. Rev. B **86**, 134524 (2012).

[5]C. Aruta, C. Schlueter, T. L. Lee, D. Di Castro, D. Innocenti, A. Tebano, J. Zegenhagen, and G. Balestrino, Phys. Rev. B **87**, 155145 (2013).

[6]M. Minola, D. Di Castro, L. Braicovich, N. B. Brookes, D. Innocenti, M. Moretti Sala, A. Tebano, G. Balestrino, and G. Ghiringhelli, Phys. Rev. B **85**, 235138 (2012).

[7]Z. Zhong, G. Koster, and Paul J. Kelly, Phys. Rev. B **85**, 121411(R) (2012)

[8]D. Samal, H. Tan, H. Molegraaf, B. Kuiper, W. Siemons, S. Bals, J. Verbeeck, G. Van Tendeloo, Y. Takamura, E. Arenholz, C. A. Jenkins, G. Rijnders, and G. Koster, Phys. Rev. Lett. **111**, 096102 (2013)

[9]S. Zhu, D. P. Norton, J. E. Chamberlain, F. Shahedipour, H. W. White, Phys. Rev. B, 54, 97 (1996); S. R. Das, P. S. Dobal, B. Sundarakannan, R. R. Das, R. S. Katiyar, J. Raman Spectrosc. **38**, 1077 (2007); M. Lavagnini, M. Baldini, A. Sacchetti, D. Di Castro, B. Delley, R. Monnier, J.-H. Chu, N. Ru, I. R. Fisher, P. Postorino, and L. Degiorgi, Phys. Rev. B **78**, 201101 (2008); N. Driza, S. Blanco-Canosa, M. Bakr, S. Soltan, M. Khalid, L. Mustafa, K. Kawashima, G. Christiani, H-U. Habermeier, G. Khaliullin, C. Ulrich, M. Le Tacon and B. Keimer, Nature Mater. **11**, 675 (2012).

[10]P. Postorino, A. Congeduti, E. Degiorgi, J.P. Itié, and P. Munsch, Phys. Rev. B **65**, 224102 (2002)

[11]C. Aruta, M. Angeloni, G. Balestrino, N. G. Boggio, P. G. Medaglia, A. Tebano, B. Davidson, M. Baldini, D. Di Castro, P. Postorino, P. Dore, A. Sidorenko, G. Allodi,

and R. De Renzi, J. Appl. Phys. **100**, 023910 (2006)

[12]D. Kan, A. Yamanaka, T. Terashima, and M. Takano, Phys. C **412-414**, 298 (2004).

[13]M. Yoshida, S. Tajima, N. Koshizuka, and S. Tanaka, S. Uchida and T. Itoh, Phys. Rev B **46**, 6505 (1992).

[14]M. V. Abrashev, A. P. Litvinchuk, M. N. Iliev, R. L. Meng, V. N. Popov, V. G. Ivanov, R. A. Chakalov, and C. Thomsen, Phys. Rev. B **59**, 4146 (1999).

[15]P. Dore, P. Postorino, A. Sacchetti, M. Baldini, R. Giambelluca, M. Angeloni, and G. Balestrino, Eur. Phys. J. B **48**, 255 (2005).

[16]V. I. Merkulov, J. R. Fox, H.-C. Li, W. Si, A. A. Sirenko, and X. X. Xi, Appl. Phys. Lett. **72**, 3291 (1998)

[17]A. A. Sirenko, I. A. Akimov, J. R. Fox, A. M. Clark, H.-C. Li, W. Si, and X. X. Xi, Phys. Rev. Lett. **82**, 4500 (1999).

[18]D. A. Tennew and X. Xi, J. Am. Ceram. Soc. **91**, 1820 (2008).

[19]G. Burns, M. K. Crawford, F. H. Dacol, E. M. McCarron III, and T. M. Shaw, Phys. Rev. B **40**, 6717 (1989).

[20]D. Di Castro, E. Cappelluti, M. Lavagnini, A. Sacchetti, A. Palenzona, M. Putti,

and P. Postorino, Phys. Rev. B **74**, 100505 (2006).

[21]S. Yoon, H. L. Liu, G. Schollerer, S. L. Cooper, P. D. Han, D. A. Payne, S.-W. Cheong, Z. Fisk, Phys. Rev. B **58**, 2795 (1998)

[22]V.B. Podobedov, A. Weber, D.B. Romero, J.P. Rice, H.D. Drew, Solid State Commun. **105**, 589 (1998).